\documentclass[useAMS,usenatbib,usegraphicx]{mn2e}
\begin{document}
\def \th {\thinspace}
\def \arcmin {\hbox{$^\prime$}}
\def \arcsec {\hbox{$^{\prime\prime}$}}
\def \chisq {$\chi ^{2}$}
\def\approxgt{\mathrel{\hbox{\rlap{\lower.55ex \hbox {$\sim$}} \kern-.3em \raise.4ex \hbox{$>$}}}}
\def\lesssim{\mathrel{\hbox{\rlap{\lower.55ex \hbox {$\sim$}} \kern-.3em \raise.4ex \hbox{$<$}}}}
\def\approxlt{\mathrel{\hbox{\rlap{\lower.55ex \hbox {$\sim$}} \kern-.3em \raise.4ex \hbox{$<$}}}}
\def \degmark {^\circ}
\def \sun {\hbox {$\odot$}}

\title[Discovery of absorption features in XB\th 1323-619]        
{Discovery of absorption features of the ADC and systematic acceleration of the X-ray burst rate in
XB\th 1323-619}

\author[Church et al.]                          
{M. J. Church$^{1 \, , \, 2}$, D. Reed $^{1}$, T. Dotani$^3$,
M. Ba\l uci\'nska-Church$^{1 \, , \, 2}$ and A. P. Smale$^4$\\
      $^1$School of Physics and Astronomy, University of Birmingham,
      Birmingham B15 2TT, UK \\
      $^2$Astronomical Observatory, Jagiellonian University,
      ul. Orla 171, 30-244 Cracow, Poland \\
      $^3$ Institute of Space and Astronautical Science, 
      Japan Aerospace Exploration Agency,
      Yoshinodai 3-1-1, Sagamihara,\\
      Kanagawa 229-8510, Japan\\
      $^4$ Office of Space Science, Astronomy and Physics Division, Code SZ,  
      NASA Headquarters, Washington, DC 20546-0001, USA}

\date{Accepted 2004 December 7. Received 2004 September 17}


\maketitle

\label{firstpage}

\begin{abstract}
We present results from analysis of the observation of the dipping, quasi-periodic bursting
low mass X-ray binary XB\th 1323-619 made with {\it XMM-Newton} in Jan., 2003. In spectral analysis
of the non-dip, non-burst EPIC PN spectrum, a number of absorption lines were discovered,
notably at 6.70 and 6.98 keV which we identify with scattering by high ionization state
ions Fe XXV and Fe XXVI. Such features have been seen in other dipping sources, but their
origin was not understood. Curve of growth analysis provided a consistent solution in which 
the line ratio was reproduced assuming collisional 
ionization with $kT$ = 45 keV, close to the electron temperature we previously
determined for the accretion disc corona (ADC) in this source. We thus propose that the absorption 
lines in the dipping LMXB are produced in the ADC. Spectral
evolution in dipping was well-described by the progressive covering model which we have
previously shown to give very good explanations of many dipping sources. We discuss the proposal
of Boirin et al. (2004b), based on analysis of the present observation, that spectral evolution in 
all of the dipping LMXB may be explained by subjecting the continuum to a highly ionized absorber. This
would require a decrease in X-ray intensity by a factor of $\sim$ 3 in dipping at energies where photoelectric
absorption is not effective (40 -- 100 keV) which previous {\it BeppoSAX} analysis of several sources
firmly rules out. In XB\th 1323-619, any decrease was less than 10$\pm$10 percent, and so the
ionized absorber proposal can be ruled out. We find a remarkable linear increase in the
rate of X-ray bursts with time over the 14-year period since 1989, and a systematic non-linear
increase in source luminosity ($L$). The linear variation of burst rate with $L$ shows that the
burst rate is proportional to mass accretion rate, and if continued implies
that the gap between bursts will become zero on January 11, 2008. In reality we expect the source
to undergo a transition from X-ray bursting to X-ray flaring which would confirm the
suggestion we previously made that flaring is unstable nuclear burning effectively consisting
of a superposition of X-ray bursts.
\end{abstract}


\section{Introduction}

XB\th 1323-619 is a member of the class of dipping low mass X-ray binaries (LMXB)
exhibiting absorption dips at the orbital period, and is also remarkable as one
of the very few X-ray burst sources in which the bursts occur regularly, but
not quite periodically. The dipping is due to absorption in the bulge in the outer accretion
disc (White \& Swank 1982; Walter et al. 1982). Spectral evolution in dipping in these sources
provides a powerful tool for constraining emission models for LMXB since not only non-dip emission 
but also several levels of dipping must be fitted. In recent years, we have shown that
the complex spectral evolution in dipping can be explained in terms of absorption
of two components: blackbody emission of the neutron star and Comptonized emission
from a very extended Accretion Disc Corona  (Church \& Balucinska-Church 1993, 1995; 
Church et al. 1997, 1998a).
The gradual removal of the non-thermal
emission in dipping shows that the ADC must be extended. Identification of the continuum
emission with the above sources was strongly confirmed by the measurement of dip ingress times,
for almost all of the dipping sources, using data from several satellites (Church \&
Ba\l uci\'nska-Church 2004). This showed that the size of the ADC varies with X-ray
luminosity between 20,000 km and 500,000 km or between 6--60 percent of the accretion disc radius.
This very large size rules out the possibility that the Comptonized emission originates in
a small central region around the compact object as has often been assumed 
(Oosterbroek et al. 2001; Sidoli et al. 2001; Gierlinski \& Done 2002).
In addition, there
are many arguments based on observational evidence that the thermal component is from
the neutron star and not the inner disc (Barnard et al. 2003); for example, the smooth 
merging of the thermal emission between X-ray bursts and the quiescent source, and the fact
that an extended ADC will cover the inner hot disc, Comptonizing the emission, so  that
disc blackbody will not be seen.
Investigation of line emission variation in dipping can indicate the source of the
line. In XB\th 1916-053, a line at 0.65 keV varies in dipping in exactly the same way as
the Comptonized emission showing that the line must originate in the ADC (Church et al.
1997). In recent years, absorption features have been discovered in several sources,
notably the galactic jet source GRS\th J1915+105 (Kotani et al. 2000), in the LMXB 
GX\th 13+1 (Ueda et al. 2001) and in the dipping LMXB, such as XB\th 1916-053
(Boirin et al. 2004a). As X-ray dips are seen in the jet source and also in GX\th 13+1,
it is clear that the absorption features are related to the high inclination of these
systems, although the exact location of the absorber has not been identified.
It thus seems likely that all dipping sources will exhibit
absorption features which can be used as diagnostics of plasma parameters in the absorbing 
region.

XB\th 1323-619 was found using {\it Uhuru} and {\it Ariel V} (Forman et al. 1978; 
Warwick et al. 1981), and dipping 
at the orbital period of 2.93 hr and X-ray bursting were discovered
using {\it Exosat} by van der Klis et al. (1985). During the {\it Exosat}
observation, the bursts repeated on a timescale of 5.30 -- 5.43 hr, approximately on
every second orbital cycle. Spectral evolution in dipping was modelled using
the ``absorbed + unabsorbed approach'' (AU) by Parmar et al. (1989), also
applied to other dipping sources (Courvoisier et al. 1986; Smale et al. 1988)
in which the non-dip spectral 
model is split into two parts for dipping. Dip spectra can be fitted if one part is
absorbed, while the other is not absorbed but requires normalization decreasing
by up to 10 times; however, this approach is difficult to justify physically.
Moreover, broadband spectra of dipping sources with {\it BeppoSAX} provide an
unambiguous test of the AU approach, since if the large decrease
in normalization necessary to fit dip spectra in the band 1 -- 10 keV is real,
then the spectra in the range 40 -- 100 keV should also exhibit a vertical shift
downwards by a large factor. It is quite clear from the {\it SAX} observations
of XB\th 1323-619 (Ba\l uci\'nska-Church et al. 1999; hereafter BC99) 
and XB\th 1916-053 (Church et al. 1998b) that there is {\it no} such decrease, and
this is strong evidence against the AU approach. Thus although the absorbed +
unabsorbed approach may {\it appear} to explain spectral evolution in instruments not extending
above 10 keV, it {\it totally fails to explain dipping in the band 0.1 -- 300 keV}.

	An explanation of spectral evolution in dipping known as the progressive covering
model was proposed more recently (e.g. Church et al. 1997) which takes into account the strong 
evidence for an extended ADC, so that the absorber gradually overlaps larger fractions of the ADC.
At any stage, a fraction of the ADC remains uncovered giving an unabsorbed
fraction of the spectrum which decreases as dipping deepens, thus explaining the need for a decreasing
normalization in the AU approach. The covered part of the ADC provides an absorbed part of the spectrum,
while the neutron star blackbody is rapidly absorbed once the point source is overlapped by absorber.
This progressive covering approach is clearly more physical, and has given extremely good fits to dip spectra
in many sources using a range of satellite observatories (Church 2001). More important, it
explains dipping in broad energy bands, such as that of {\it BeppoSAX}, i.e. in the region
{\it 10 -- 100 keV}, which the absorbed and unabsorbed approach is unable to do, which
therefore must be rejected as incorrect.

XB\th 1323-619 was observed by us with {\it BeppoSAX} 1997, August 22 
(BC99) and using {\it Rossi-XTE} on 1997, April 25
(Barnard et al. 2001). The {\it BeppoSAX} observation allowed a detailed study of the
X-ray spectrum, and of dipping and bursting. The broadband spectrum in the range 1--200 keV
allowed accurate determination of emission parameters, and the relatively high cut-off
energy of 44 keV of the Comptonized emission, ten times higher than in many LMXB,
proves a high electron temperature in the ADC. The neutron star blackbody emission
had kT = 1.77$\pm$0.25 keV contributing $\sim$15 percent of the 2 -- 10 keV flux. 
\begin{figure*}                                                         
\includegraphics[width=100mm,height=170mm,angle=270]{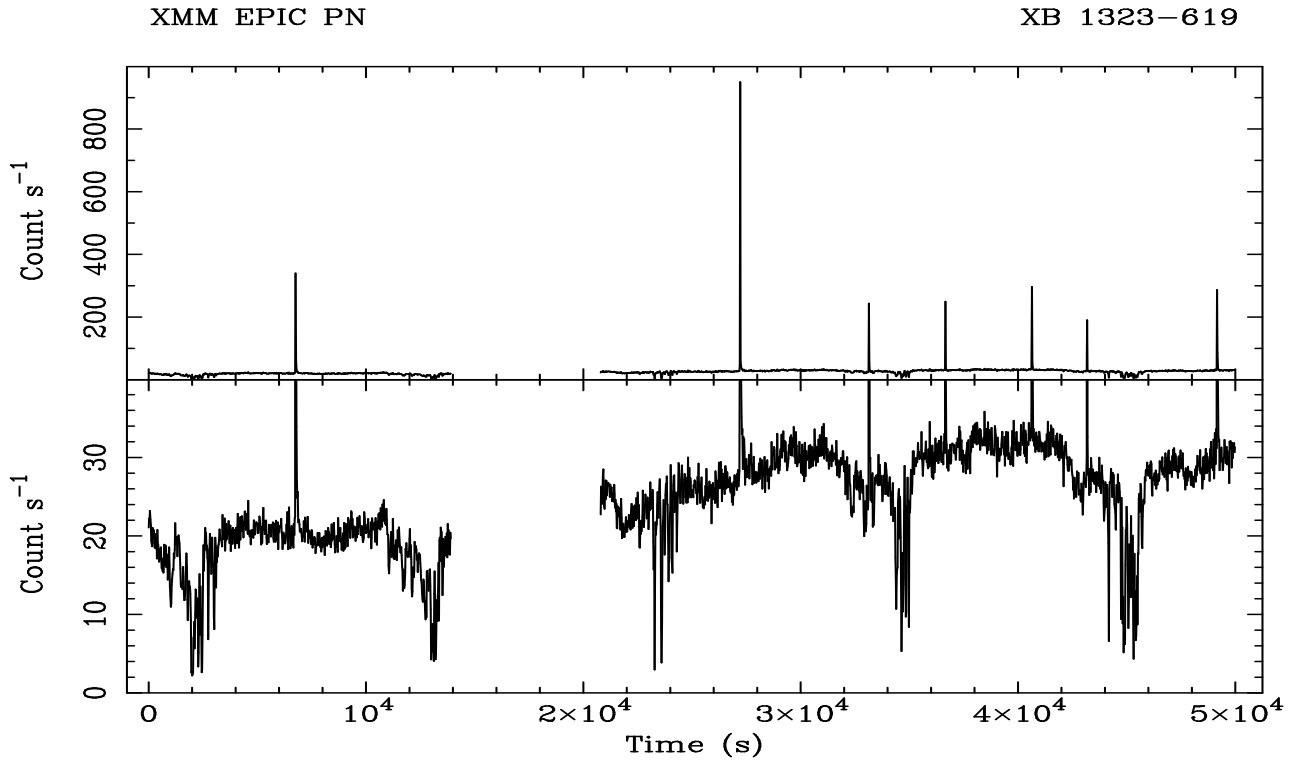}               
\caption{EPIC PN lightcurve of XB\th 1323-619 in the band 0.5--10.0 keV: lower panel shows
the X-ray dipping; the upper panel shows the X-ray bursts more clearly with burst heights
corrected (see text)}
\label{}
\end{figure*}
%
Spectral analysis of a burst that took place during a dip interval exhibited a surprising lack of
absorption shown to be consistent with ionization of the absorbing bulge by the X-ray
burst (BC99), showing that during the burst {\it all}
of the accretion disc must be highly ionized. The {\it RXTE} observation giving
100 times improved count statistics could not be fitted without including the effects
of dust scattering, and the optical depth to dust scattering was obtained from the
radial intensity profile of the {\it BeppoSAX} data (Barnard et al. 2001). The {\it RXTE}
observation also revealed a broad iron emission line at 6.43$\pm$0.21 keV with 
equivalent width (EW) 110$\pm$55 eV.

During the above observations, it was seen that the burst rate was increasing; for example
in the {\it BeppoSAX} observation the bursts were repeating every 2.40 -- 2.57 hr. Although
this might be expected to be due to changes in mass accretion rate, i.e. source
luminosity, there was at this stage no apparent systematic change in the luminosity.
The source luminosity (1 -- 10 keV) was $\sim 2\times 10^{36}$ erg s$^{-1}$ 
entirely consistent with the observation of X-ray bursts which are seen in
lower luminosity sources, but not in brighter LMXB closer to the Eddington limit which
exhibit X-ray flaring. 

In the present paper, we present analysis results for our observation of the source
with {\it XMM-Newton} in which several absorption features were discovered. In
addition, we find that the burst rate has increased substantially, and we can
now show that this is very well correlated over a period of 14 years with a 
systematic increase in the source luminosity, presumably reflecting an increase
in mass accretion rate from the Companion, the reason for which is not clear.

\section{Observations and analysis}

We observed XB\th 1323-619 on 2003 Jan 29 from 09:03 -- 22:47 UT for a total observation of 50 ks 
containing a usable exposure of 43 ks. The European Photon Imaging Cameras (EPIC) consist of two MOS 
and one PN spectrometer. Because the PN instrument is effectively two times more sensitive than
the MOS as it diverts no flux to the grating instruments, we present PN results here. Raw data were assembled
into events files which were then used to provide lightcurves, images and spectra using the
Science Analysis Software SAS, version 5.4.1. Hot and flickering pixels
and the neighbouring pixels of these were removed during the processing. Standard processing
was used so that CCD events in single or double pixels produced by a single photon were
accepted and other events rejected. There was no
pulse pileup effect as the count rate even at the peak of the largest X-ray burst was
well below that at which pileup is first seen. Barycentric correction was carried out, and finally, 
during a period of high particle background from 13.9--20.8 ks, science data were not accumulated.

	The PN detectors were operated in timing mode in which only a single CCD containing the
source is used, and image data were not available since the data are projected onto one dimension.
The field of view of the single chip is 199$\times$ 64 pixel, where 1 pixel is 4.1\arcsec\,  i.e.
$13.\arcmin 6 \times 4.\arcmin4$. We extracted source events from a column 70\arcsec\ wide centred
on the source position, and obtained background from two columns each 35\arcsec wide at the edges
of the strip, each $\sim$115\arcsec\ from the source. 
The field of view of the detector was such that
the X-ray pulsar discovered during the {\it BeppoSAX} observation of XB\th 1323-619 
(Angelini et al. 1999) and located 17\arcmin away, was not included and could not contaminate the 
data. The latest instrument response file for the PN (epn\_ti40\_sdY9.rsp) released in January, 2003, 
was used.
\begin{figure}                                                         
\includegraphics[width=70mm,height=84mm,angle=270]{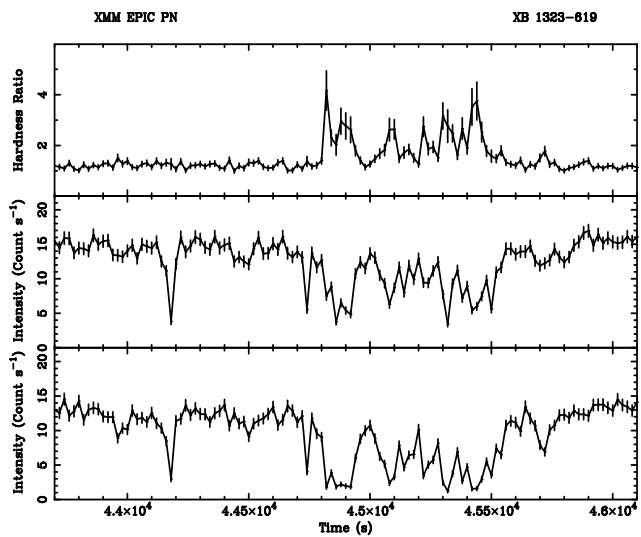}                
\caption{Detailed view of the 5th dip in two energy bands: centre panel: 0.5--3.5 keV, lower panel:
3.5--10 keV together with the hardness ratio formed from dividing these (top panel). 
The marked spectral hardening in dips can be seen, except for the unusual events
 at the start of dipping (see text)}
\label{}
\end{figure}

\section{Results}

	The background-subtracted X-ray lightcurve obtained from the EPIC PN in the energy range 
0.5 -- 10 keV is shown in Fig. 1 with 20 s binning (lower panel). 5 intervals of dipping can 
be seen, and the 7 X-ray bursts are displayed separately in the upper panel.
Two of the bursts occur in the wings of X-ray dips, although no bursts
coincide with deep dipping. With 20 s binning, bursts do not have their true height and so
we examined each burst individually with shorter binning, decreasing the bin size
until at a few seconds, there was no change in burst height. The bursts are
shown for clarity separately in the upper panel of Fig. 1 with 
corrected heights. It is striking that one burst is very much more intense 
than the others with peak intensity 35 times that of the persistent emission, about 3 times
larger than the other bursts. The two bursts coinciding with the wings of X-ray dips
are slightly reduced in intensity compared with the other bursts. The mean interval between 
bursts in the second half of the observation containing 6 bursts is 58 min. However, 
in the first part of the observation only one burst is seen when 3 might have been
expected, and there is more variation in burst height, suggesting that the bursting is 
rather more erratic than previously, e.g. in {\it BeppoSAX}, possibly associated with the
increased source luminosity (Sect. 3.3).

Dipping is deep, the intensity falling by 85 percent of the non-dip value and exhibits a high degree of
variability. The data were searched for periodicities to obtain a measurement of the orbital
period from the dip recurrence using the {\sc ftool} {\sc powspec} and a period of 2.980$\pm$0.053 hr obtained
with PN data, and 2.961$\pm$0.048 hr with MOS. The average of these is 2.97$\pm$0.05 hr which
is consistent with the period obtained from {\it BeppoSAX} of 2.938$\pm$0.020 hr (BC99).
In Fig. 2 we show an expanded view of the variability in the fifth 
dip, together with the associated increase in hardness ratio, defined as the intensity in 
the band 3.5--10 keV divided by that in the band 0.5--3.5 keV. The deep dipping within the dip
envelope displays increases in hardness as expected; however the first two dip events
have no
associated hardening, which is very unusual. We examined the lightcurves
in the bands 0.5--3.5 and 3.5--10.0 keV and found these dip events to have exactly equal 
depth. The only exception previously found to the photoelectric absorption that produces 
X-ray dips, is the occurrence of electron scattering in the ``shoulders of dipping''
in the brighter dip sources X\th 1624-490 and XB\th 1254-690 (Smale, Church \&
Ba\l uci\'nska-Church
\begin{figure}                                                              
\includegraphics[width=60mm,height=84mm,angle=270]{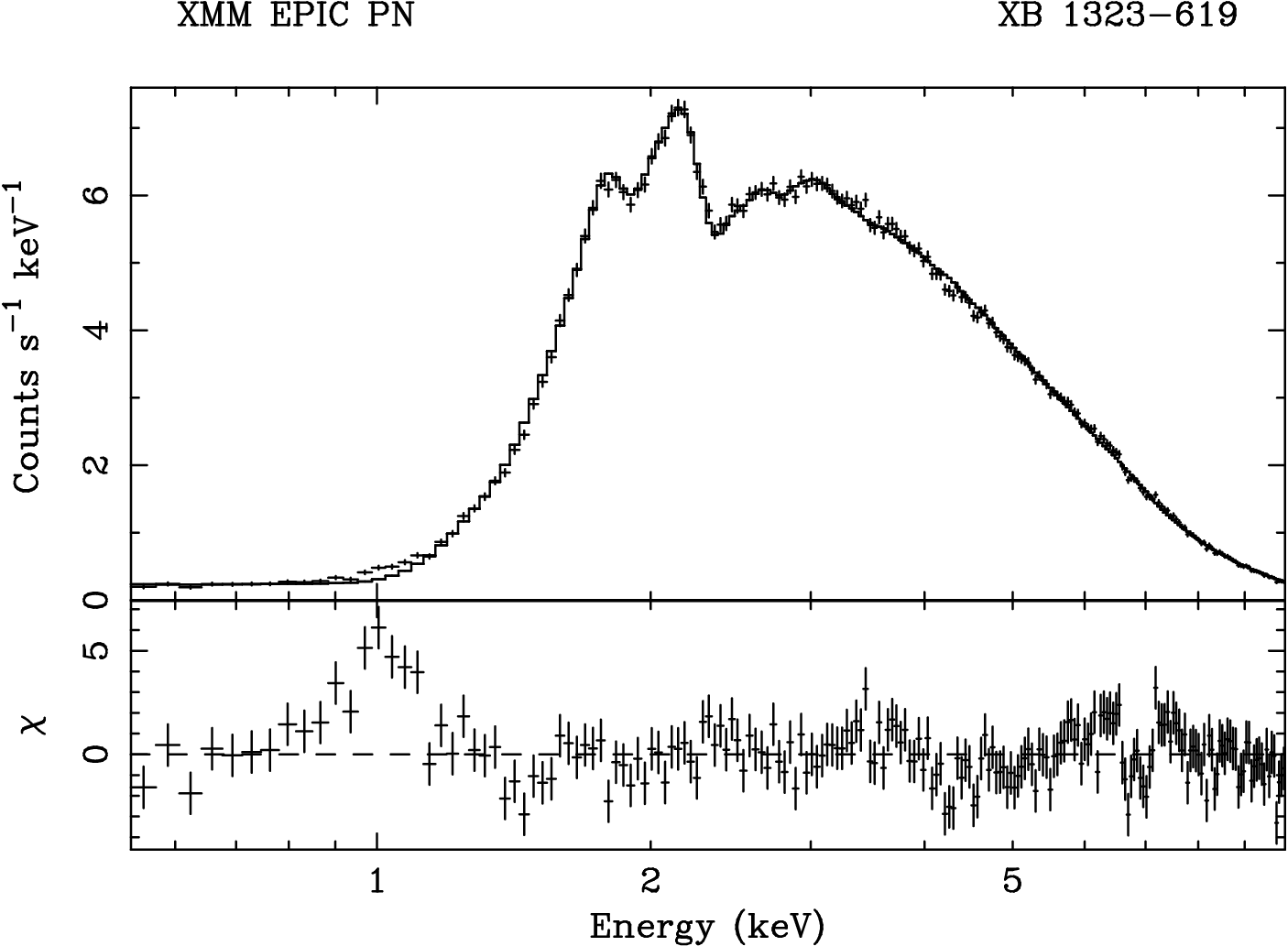}                    
\includegraphics[width=20mm,height=84mm,angle=270]{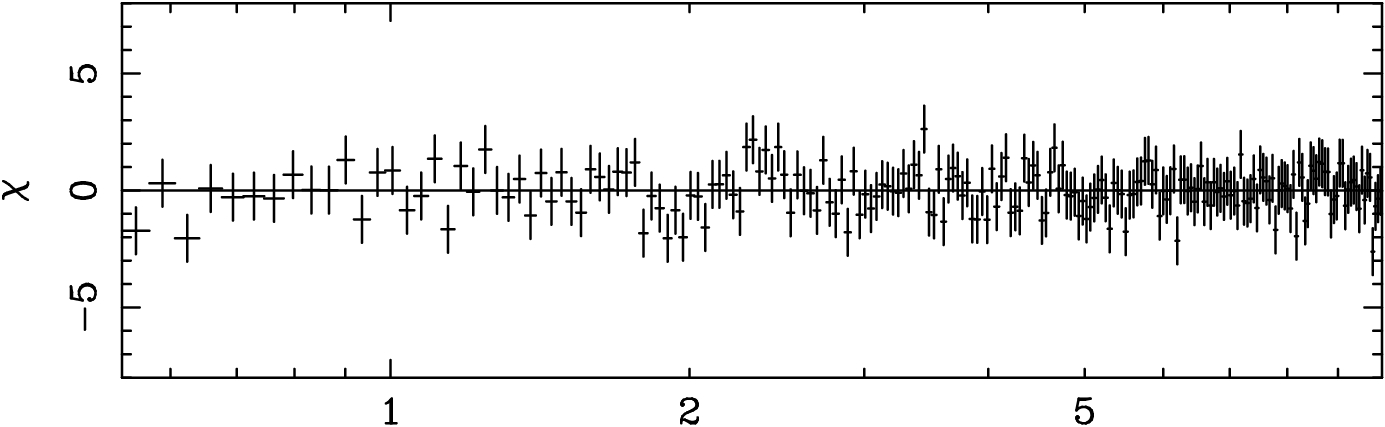}                     
\caption{Spectral fit of the non-dip, non-burst spectrum. The upper panel shows the folded data
and residuals before addition of lines to the spectral model, while the lower panel shows the
residuals after the lines were added}
\label{}
\end{figure}
20001, 2002), i.e. when the lightcurves
display dipping at about the 10 percent deep level before deep dipping starts, which we have explained
as electron scattering in the outer highly ionized layers of the absorber. This is not seen
in conjunction with fast variability in dipping which is due to absorption of the point-like neutron 
star emission. Thus it appears that this event is caused by the neutron star emission being either 
fully absorbed by very dense matter or reduced by electron scattering in fully ionized material
probably on the outer disc.

In the band 0.5--2.0 keV, dipping is
100 percent deep, showing firstly that the absorber subtends a larger angle at the neutron star than
the extended emission component. This deep dipping is also initially surprising in view of the dust 
scattering previously found with this source. For sources with higher Galactic column densities, scattering takes 
place both out of and into the beam. Because of the delay involved in the scattering process,
a fraction of the non-dip emission is added to the dip intensity  reducing the
observed depth of dipping. Thus in the observations of XB\th 1323-619 with {\it BeppoSAX}
(BC99) and {\it RXTE} (Barnard et al. 2001) the depth of dipping did not reach 100 percent
although other factors also contributed to this.
In the case of {\it BeppoSAX}, data were extracted from circular regions around the source of
radius 4$\arcmin$ for the MECS, while in {\it RXTE}, no selection was possible as the PCA is
not imaging. It was concluded that the dust scattering halo contributed a few percent in
the case of {\it RXTE}, this having been derived from fitting the radial intensity profile in
{\it BeppoSAX}, so that dipping could never be 100 percent deep. We investigated dust scattering
in the present {\it XMM} observation by re-binning the MOS image into radial bins in the range
0.5 to 300 arcseconds, and comparison of this with the point spread function reveals a clear
excess above the PSF at $r$ $>$ 40$\arcsec$ which 
\begin{figure}                                                         
\includegraphics[width=60mm,height=84mm,angle=270]{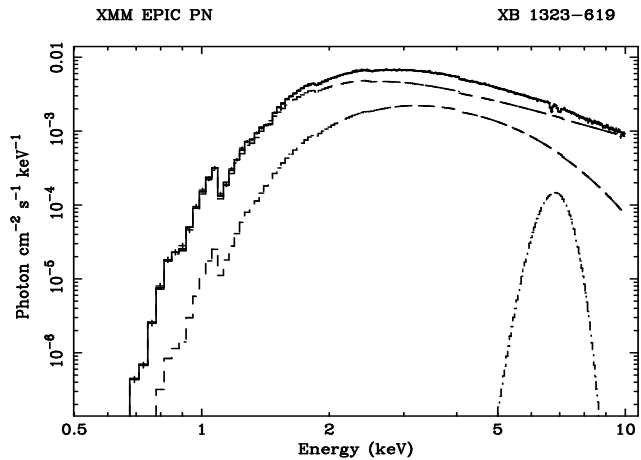}                 
\caption{The unfolded non-dip, non-burst spectrum with best-fit model consisting of continuum
plus line and edge features}
\label{}
\end{figure}
can be identified with the dust scattering
halo. However, data selected for analysis were obtained from a region of size 35$\arcsec$ and hence
dust scattering does not prevent dipping becoming 100 percent deep.

\subsection{The EPIC PN spectrum of persistent emission}

Non-dip, non-burst intervals of data were selected from the lightcurve of XB\th 1323-619,
and to obtain the best spectrum for detection of lines, the whole observation 
was used (excluding bursts) although there was a 50 percent increase in luminosity during the observation.
In analysis of dipping, only data from the second half of the observation were used
where the intensity was stable to avoid mixing data of different intensities.
In our previous work, we have shown that the continuum spectra of the dipping LMXB, and
indeed of LMXB in general, can be best-fitted by a two-component model consisting of
blackbody from the neutron star and Comptonized emission from an extended ADC (Sect. 1).
In the high quality, broadband {\it BeppoSAX} data on XB\th 1323-619 this was also the case
(BC99). For this reason, we apply the above continuum
model to the above data as the evidence for it is so strong. We note that in analysis
of {\it XMM} data on other dipping sources, Parmar and co-workers (e.g. Boirin
et al. 2004a) choose a model containing a disc blackbody component although there is
little evidence to support this.

Primitive channels were binned together such that, at any energy,
the energy resolution was oversampled by a factor of three; in addition the data were grouped
to a minimum of 20 counts per bin so that the $\chi^2$ statistic could be used. 
%
\begin{table*}
\begin{center}
\caption{Line and possible edge features detected in the non-dip spectrum of XB\th 1323-619.
In the case of the emission line, the measured energy may have been affected by the
neighbouring absorption lines, and identification of the transition is uncertain}
\begin{tabular}{|c|c|c|r|c|c|c|}
\hline
 E (keV) & Em/Abs & EW (eV) & F-test prob. &Transition & Energy (keV)\\
\hline \hline
1.09$\pm$0.01 & edge & \dots & $3\times 10^{-23}$ & Ne VII/VIII, Fe XIII/XIV & 1.08,1.13,1.08,1.13\\
1.46$\pm$0.03 & A  & -15   & 1$\times 10^{-4}$ & Mg XII & 1.48 \\
4.05$\pm$0.08 & edge   & \dots & $1\times 10^{-6}$ &Ca XX K$\alpha$ & 4.03\\
6.70$\pm$0.03 & A  & -22 & $2\times 10^{-7}$ &Fe XXV K$\alpha$ & 6.68\\
6.98$\pm$0.04 & A  & -27 & $1\times 10^{-6}$ & Fe XXVI K$\alpha$ & 6.95 \\
6.82$\pm$0.11 & E  & 92  & $3\times 10^{-6}$ & Fe XXV/XXVI & 6.68/6.95 \\
\hline
\end{tabular}
\end{center}
\end{table*}
\begin{figure}                                                      
\begin{center}                                                      
\includegraphics[width=32mm,height=84mm,angle=270]{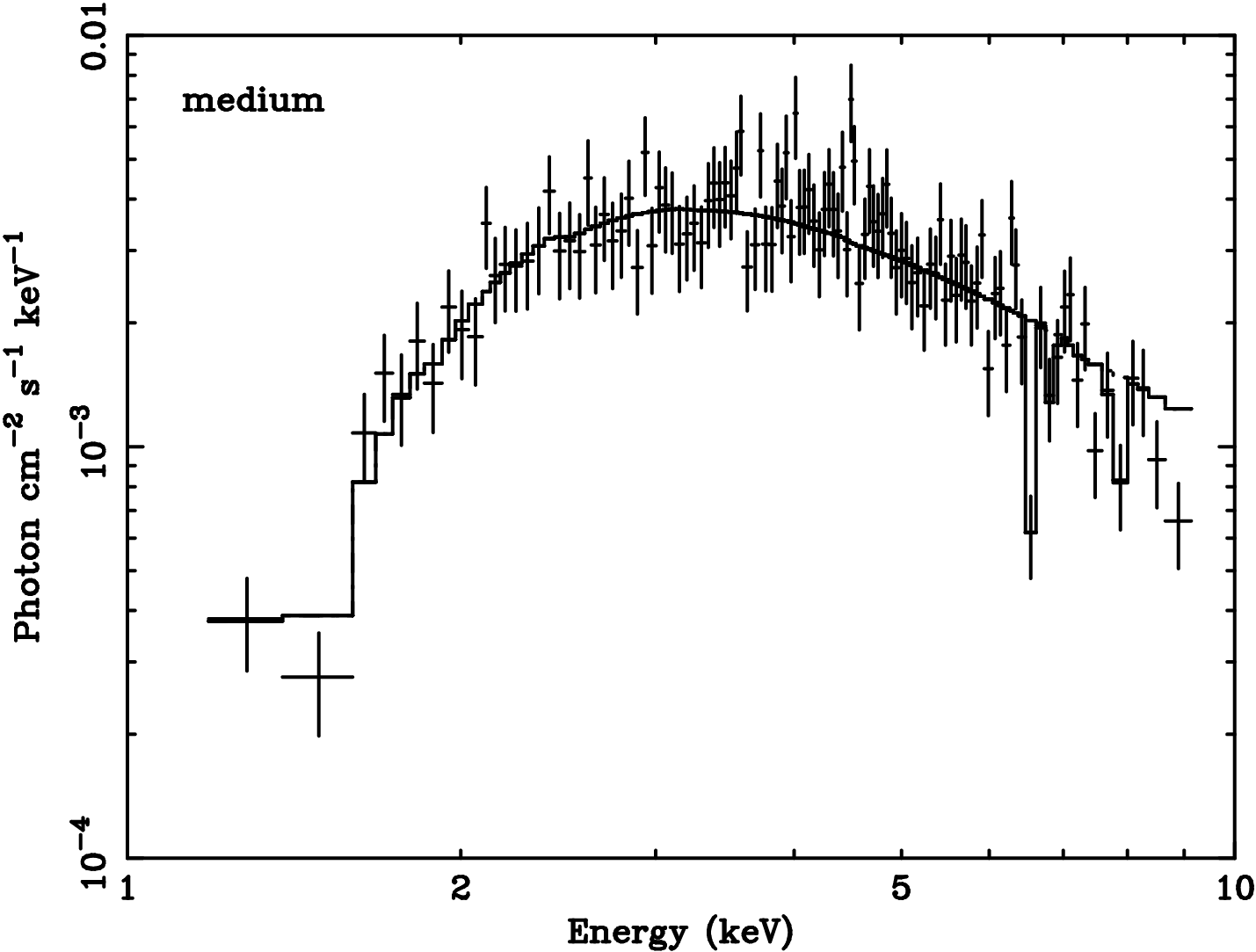}
\includegraphics[width=32mm,height=84mm,angle=270]{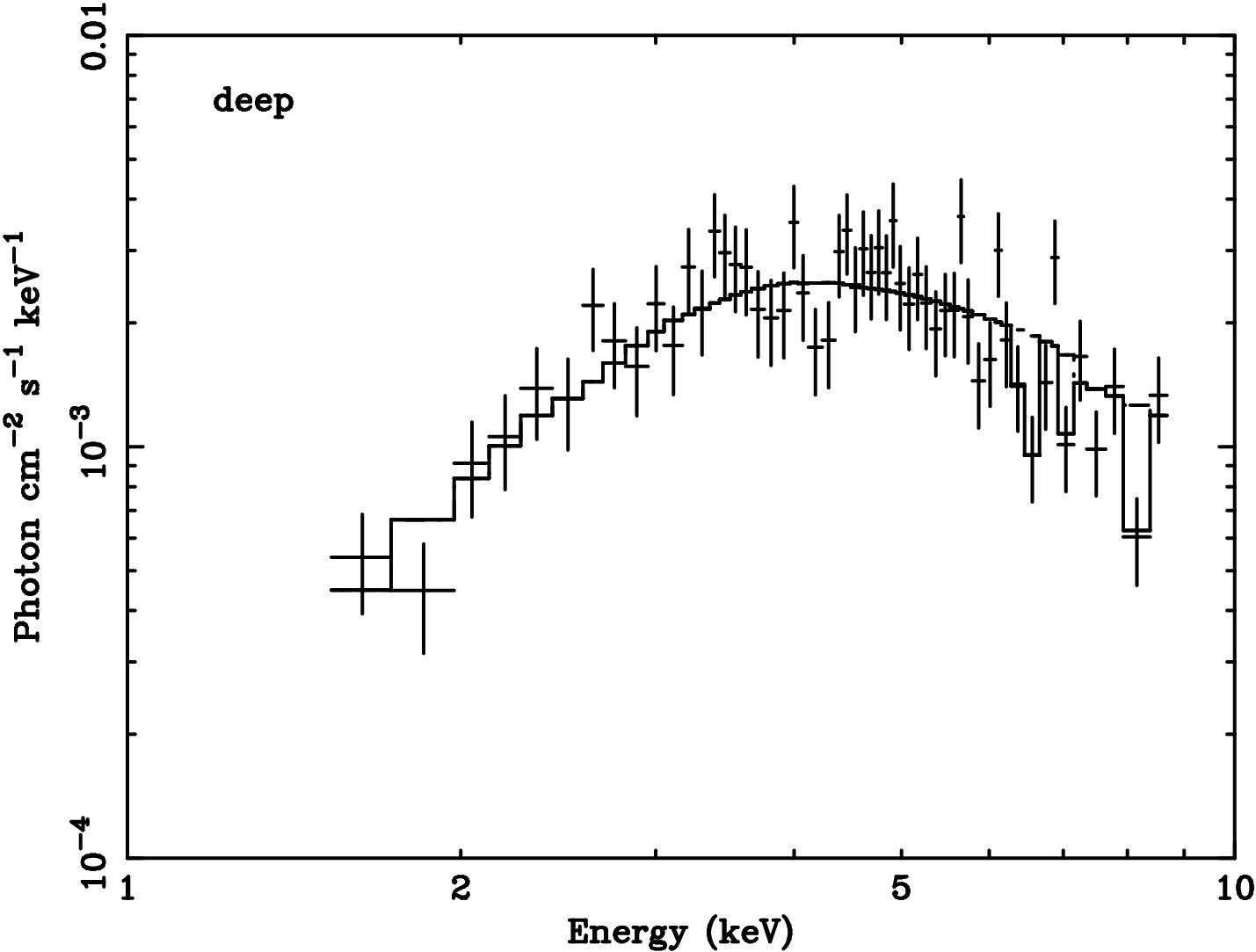}
\includegraphics[width=32mm,height=84mm,angle=270]{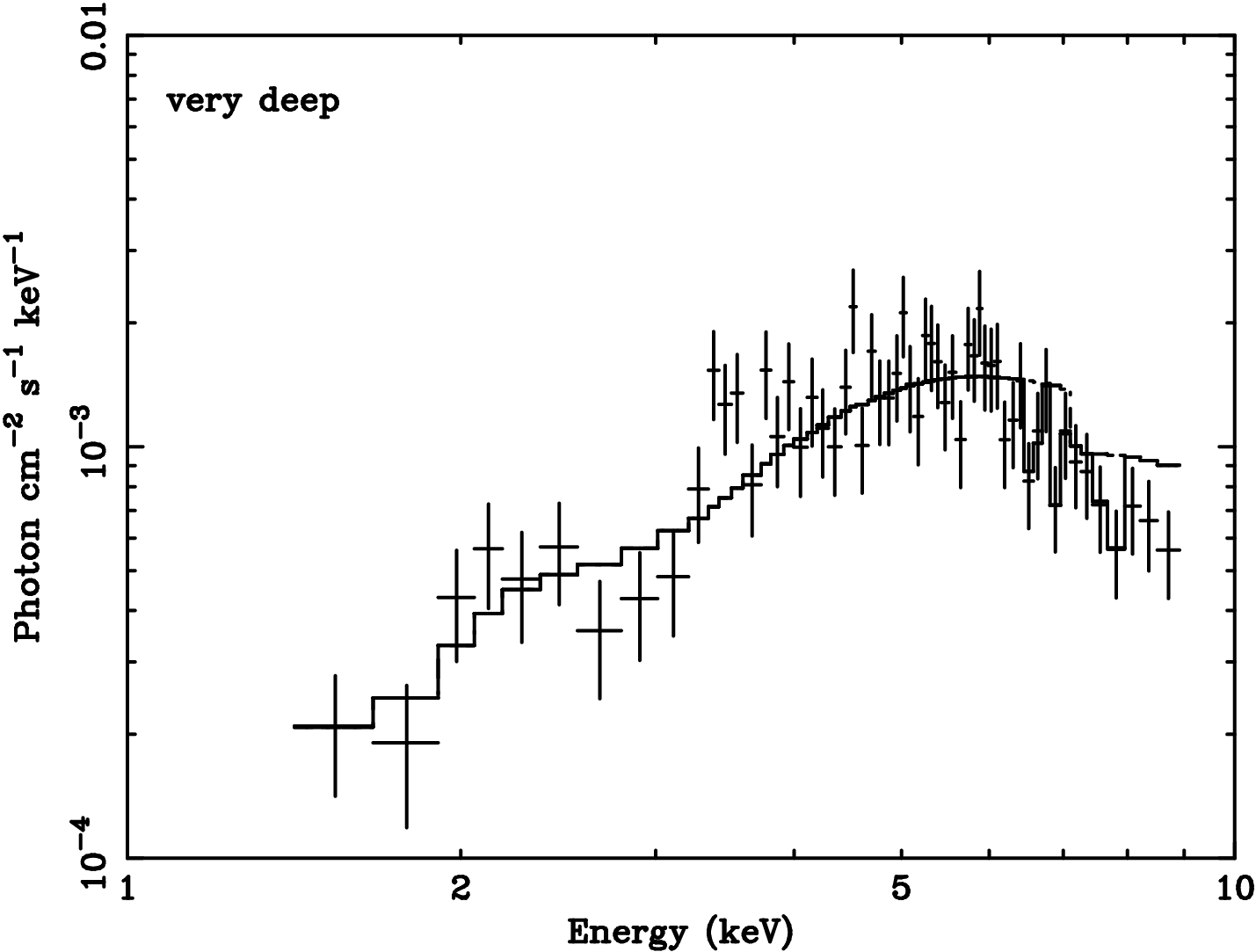}
\caption{Spectra of three levels of dip emission showing that the progressive covering
model provides very good fits to the continuum at all levels.}
\end{center}
\end{figure}
Systematic errors of 2 percent
were added. Data were fitted in the band 0.52--10.0 keV, so avoiding a point at 0.5 keV
which appeared anomalous.
Fig. 3 shows the folded spectrum of the persistent emission fitted by a continuum model
only, together with the residuals, from which it is immediately clear
that various absorption and emission features were present. Firstly,
there is a broad feature at $\sim$1 keV. This can be fitted either by an emission line
or by an absorption edge. A comparison of $\chi^2$ values can best be made when all
other absorption and emission features are added, and at that stage, there is no
significant difference in goodness of fit of emission line or edge.
If an emission line is used, the energy
is 0.80 keV corresponding to Fe XXVII or O VIII although the line is broad and very
strong (EW $\sim$ 4 keV), and without apparent structure. The energy appears higher in Fig. 3 
because the line is modified by absorption. An edge at $\sim$1.1 keV could
consist of Ne VII (1.078 keV), Ne VIII (1.125 keV), Fe XIII (1.081 keV) or Fe XIV (1.125 keV).
We choose to adopt the edge model, because of the possibly unreal width and strength
an emission line would need. However, there is evidence from dip spectra (below) 
\begin{figure}                                                      
\begin{center}                                                      
\includegraphics[width=62mm,height=84mm,angle=270]{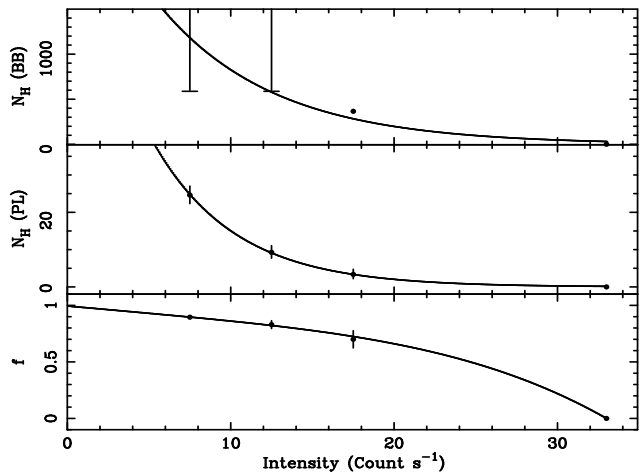}
\caption{Best-fit spectral fitting results : variation of blackbody column
density, power law column density and power law partial covering fraction
{\it f} with intensity. $\rm {N_H}$ is in units of $\rm {10^{22}}$ H atom
$\rm {cm^{-2}}$}
\end{center}
\end{figure}
marginally
against the edge model: when we added the two lowest dip levels together to obtain a
spectrum extending down to 0.5 keV, but with only two bins spanning 0.6--1.5 keV.
Although the spectrum at 1 keV was of poor quality, there was no sign of any feature 
at 1 keV. While we expect an emission line to be absorbed in dipping, an edge would 
probably not be removed. Thus the edge nature of the 1 keV feature cannot be regarded as quite definite.

Using the edge model, a number of absorption features can be seen in the residuals,
notably the lines at 6.70 keV and 6.98 keV (Fe XXV and XXVI). There was also a weak feature at $\sim$1.46 keV
which might correspond to Mg XII, but an EW of a few eV does not constitute a significant
detection. There was also some sign of weak, narrow absorption at 7.8 keV corresponding
to Ni XXVII. A feature at 4.0--4.6 keV could be fitted as an absorption line or as an edge.
If fitted as a line, the energy obtained of $\sim$4.6 keV
does not correspond to a known line, but an edge at $\sim$4.0 keV may be Ca XX.
Starting from the continuum-only fit, lines were
added to the spectral model sequentially, until the final model was obtained having the lines
shown in Table 1. The best-fit is shown in Fig. 4 as an unfolded spectrum, and the
remaining small residuals shown in Fig. 3 (lower panel). From these, it can be seen that
the broad positive residuals between 6--8 keV in the continuum model were partly a result
of the absorption features not being modelled; in the final fit, this resolves itself
into a single emission line, although the measured energy of 6.8 keV was probably
distorted by the neighbouring absorption lines, making 
\begin{figure}                                                  
\begin{center}                                                  
\includegraphics[width=70mm,height=84mm,angle=270]{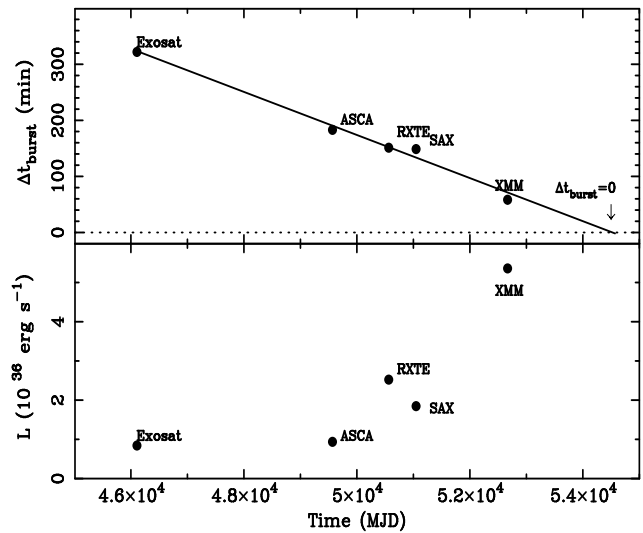}
\caption{Top panel: the long-term variation of the mean interval between bursts, fitted by a
linear model; $\Delta$t errors, typically $\pm$5 min, are too small to be seen.
Lower panel: the 1--10 keV source luminosities for each of the observations}
\end{center}
\end{figure}
identification uncertain.
The interpretation of these features is discussed in Sect. 4. The best-fit values of 
the continuum parameters were $kT$ = 1.09$^{+0.07}_{-0.02}$ keV 
for the neutron star blackbody, power law photon index $\Gamma$ = 1.69$\pm$0.08, and 
the column density $N_{\rm H}$ was $\rm{2.42\pm 0.14\times 10^{22}}$ atom cm$^{-2}$.
The column density is somewhat larger than the Stark et al. (1992) value of 
$\sim 1.4\times 10^{22}$ atom cm$^{-2}$.

\subsection{Spectral evolution in dipping}

Dip spectra were produced by selecting in intensity bands from the second half of the 
observation. X-ray bursts were removed, and intensity bands defined 
of 5--10, 10--15, 15--20 count s$^{-1}$ for deep, medium and shallow dipping, and also
the non-dip spectrum at 30--36 count s$^{-1}$. The non-dip spectrum 
contained only data from times later than 30 ks in the observation, to exclude earlier
data at lower intensity. This spectrum was fitted to obtain the emission parameters:
$N_{\rm H}$, $kT$, $\Gamma$ and the normalizations, and these were frozen in fitting the
dip spectra as only absorption parameters can change. The same line and edge features
as found above were included in fitting each spectrum.
The spectral model tried in fitting had the form
{\sc ag*(ab*bb+pcf*pl)}, where {\sc ag} is Galactic absorption, {\sc bb} is the neutron
star blackbody emission, {\sc ab} the absorption of the blackbody 
and {\sc pl} is the Comptonized emission of the extended ADC.
The point-like blackbody is allowed to be covered rapidly once the envelope of the
absorber passes across the neutron star, but the ADC is subject to progressive 
covering modelled as {\sc pcf}: a partial covering fraction as the absorber progressively
overlaps more of the ADC. This ``progressive covering'' model (Sect. 1) has previously provided 
very good fits to observations of the dipping sources, including XB\th 1323-619 (Church et al. 1997;
BC99).

Good fits were obtained for the dip spectra with the emission parameters frozen at the non-dip
values, with $\chi^2$/d.o.f. = 104/101, 46/40 and 57/45, for medium, deep and very deep dipping,
and these fits are shown in Fig. 5 where simple fits are added to the data.
\begin{figure}                                                     
\begin{center}                                                          
\includegraphics[width=70mm,height=84mm,angle=270]{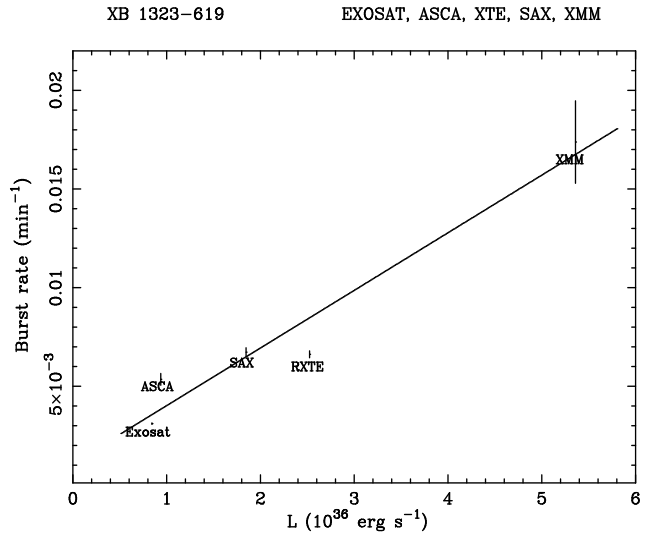}
\caption{The data of Fig. 7 re-plotted as 1/$\Delta$ t, i.e. as burst repetition rate as a function
of $L^{1-10}$ with best-fit simple linear model}
\end{center}
\end{figure}
Clearly the two deeper dip spectra are not of the highest
quality because of the lack of counts. The variation of the absorption parameters within the
progressive covering model 
are shown in Fig. 6. It can be seen that the covering fraction rises smoothly from zero
to 95 percent in the deep dip spectrum, and that the column density is much higher for the
neutron star blackbody emission than for the extended emission, and only lower limits
can be given for the lower two dip levels. The reason that this column density is higher
is that
the neutron star effectively allows measurement of $N_{\rm H}$ along a line through
the centre of the absorber, whereas the extended emission has $N_{\rm H}$ determined
as an integration over the absorber including lower density outer regions. It is quite
clear that the above absorber modelling gives good fits, providing further evidence
for the progressive covering model which gives
a clear physical explanation of spectral evolution in dipping, and thus evidence
against the alternative ``absorbed plus unabsorbed'' approach.

The absorption lines of the non-dip spectrum were found to become more pronounced in dipping, 
in particular the Fe XXV and XXVI lines. In addition, it was striking that an
additional absorption line at $\sim$7.8 keV corresponding to Ni XXVII K$_{\alpha}$ was clearly
detected, which although just seen in the non-dip spectrum, did not constitute a significant detection.
The equivalent width of the Fe XXV line in the deep dip spectrum was $\sim$100 eV, 
that is, five times larger than in the non-dip spectrum. The increase in EW reflects an increase
of integrated flux in the line in dipping by a factor of 2.5 compared with non-dip, while the continuum
spectral flux density decreases by a factor of two, thus there is clearly additional absorption
in the line. The relatively poor statistics of the dip spectra tended to mask systematic trends in the
line parameters; however, the Fe XXV line energy in the dip spectra was 6.55$\pm$0.1 keV.
%
This is consistent with a decrease of the ionization state
of the additional absorber involved in forming absorption lines in dipping.
Finally, the strong emission line at $\sim$6.8 keV was not present in the dip spectra
implying that this was produced in the ADC so that in dipping
it is removed as the ADC is overlapped by absorber. 

\subsection{14-year acceleration increase of rate of X-ray bursts}

We next present results on the clear systematic change of X-ray burst rate over the 14-year interval
since the {\it Exosat} observation. For the {\it RXTE}, {\it ASCA} and {\it SAX} observations,
we use our previously published burst rates (Barnard et al. 2001;  BC99); for {\it Exosat} we
also use the published rate (Parmar et al. 1989). In the case of the {\it Exosat}, {\it RXTE} and {\it SAX}
observations, the bursting clearly displayed the regularity for which the source is known, i.e. most bursts
repeated regularly. However, in obtaining the mean burst separation, we did exclude bursts
following double bursts which cause a glitch in the burst repetition rate. In the {\it XMM}
observation, it appears that the degree of regularity has decreased presumably associated with
the luminosity increase. Thus Fig. 1 shows that in the second half of the observation, the
bursts tend to be regular, but this is not seen in the first half. Consequently, we make the best
estimate possible from the second half. This value of 58.4$\pm$7 min between bursts is remarkably 
shorter than in previous observations of this source. In {\it Exosat} (1985, February) the
separation was 5.30 -- 5.43 hr (Parmar et al. 1989), in {\it ASCA} (1994, August) it was 3.05 hr
(BC99); {\it XTE} (1997, April) it 2.45 -- 2.59 hr (Barnard et al. 2001) and in 
{\it BeppoSAX} (1997, August) it was 2.40 -- 2.57 hr (BC99).
Thus it is very clear that the rate of bursting has been increasing systematically
since 1985 and probably before that. 

To investigate the dependence of burst rate with source luminosity, we have compiled the values
of source luminosity from our spectral analysis of the {\it ASCA}, {\it RXTE} and {\it BeppoSAX}
observations. In addition we obtained the {\it Exosat} ME spectrum and instrument response from 
the HEASARC archive and carried out spectral fitting of this using the same two-component model
used with the other observations (see Sect. 3.1). In all cases we derived the flux and luminosity
in the 1 -- 10 keV band, assuming a distance of 10 kpc, to avoid extrapolating too far from
the actual instrumental energy ranges. The results are shown in Fig. 5 in the form of
burst recurrence time $\Delta$t as a function of MJD (upper panel), and source luminosity $L$ (lower panel).
The linear variation of burst rate is remarkable, and is presumably due to the
systematic increase of luminosity over the 14 year period covered. It should be noted that 
the data from {\it Exosat}, {\it RXTE} and {\it SAX} by themselves prove the change in burst
rate; the {\it XMM} data are clearly consistent with this trend, even if the regularity of
the bursting had decreased. Least squares fitting
of the data in Fig. 5 produced a best-fit linear slope of -0.03847 min MJD$^{-1}$
such that if we assume the linear behaviour continues, the gap between bursts would become zero
on MJD 54476, i.e on 2008, Jan 11. In reality, we do not expect this to happen, since only
low luminosity sources exhibit X-ray bursts while brighter sources exhibit flaring; thus we expect 
the source to undergo a gradual transition to flaring before 2008 as the luminosity increases
from 10$^{37}$ to 10$^{38}$ erg s$^{-1}$.

We also plotted 1/$\Delta$t as a function of $L$ to test the hypothesis that the burst rate
is increasing because of the luminosity increase: for a given mass accretion rate $\dot M$,
the gap $\Delta$t is determined by the time to accumulate sufficient mass on the neutron star
surface for a nuclear event. Thus 
%
\begin{table}
\begin{center}
\caption{Column densities of absorbing ions for the non-dip spectrum from curve of growth analysis
shown as logarithms. Also shown are implied values of 
$kT_{\rm therm}$ if the ion distributions are due to collisional ionization,
and values of the ionization parameter $\xi$ for ions produced by photo-ionization.
For each ionization process, the hydrogen column density of the plasma is given
in non-logarithmic form}
\begin{tabular}{|l|c|c|c|}
\hline
specie & $kT_{\rm CG}$ = 100 keV &  $kT_{\rm CG}$=45 keV & $kT_{\rm CG}$=1 keV\\
\hline \hline
Fe XXV K$_{\alpha}$        &  17.6$\pm$0.1  & 17.8$\pm$0.1 &   18.7$\pm$0.2\\
Fe XXVI K$_{\alpha}$       &  18.0$\pm$0.1  & 18.8$\pm$0.1 &   19.95$\pm$0.1\\
thermal ionization&&&\\
$N_{\rm H}$ (atom cm$^{-2}$)  & 9.9$\times 10^{22}$ & 1.7$\times 10^{24}$ & 2.7$\times 10^{25}$\\
$kT_{\rm therm}$ (keV)                 &  14            &  42          &   100\\  
photo-ionization&&&\\
$N_{\rm H}$ (atom cm$^{-2}$)  &  4.3$\times 10^{22}$& 4.0$\times 10^{23}$& 8.0$\times 10^{25}$\\
$\xi$  (erg cm s$^{-1}$)      &  10$^{3.4}$    & 10$^{3.6}$   &   10$^{4.3}$\\
\hline
\end{tabular}
\end{center}
\end{table}
with increasing $L$ and $\dot M$, 1/$\Delta$t should be 
proportional to $L$. Fig. 6 shows that this is indeed the case, with a simple linear fit
added to the data. This source is remarkable in having burst rate $\propto$ $\dot M$
as also found recently for the other quasi-periodic burster GS\th 1826-238 by Galloway et
al. (2004). Although this may be expected theoretically, the majority of burst sources
have a burst rate that {\it decreases} with increasing $\dot M$ (e.g. Strohmayer \&
Bildsten 2003), perhaps because the burning area increases so that the mass flow per
unit area may decrease.

\section{Discussion}

Fitting the spectra of XB\th 1323-619 has shown that both the emission model used,
and the progressive covering absorption model provide very good fits. 
Thus the non-dip continuum spectrum is
well-described by assuming blackbody emission from the neutron star plus Comptonized
emission from the extended accretion disc corona. Fig. 5 shows that the absorption model 
provides excellent fits to several levels of dipping, this model assuming that the extended
ADC is covered gradually by the extended absorber, so that at every stage of dipping, part of
the ADC is covered by absorber and part is not. This progressive covering explanation
of spectral evolution in dipping has been shown to explain dipping in many dipping sources,
and crucially, is able to explain dipping in broadband X-ray observatories such as {\it SAX},
i.e. it predicts that {\it no} change will take place in the range above 40 keV where there
is no photoelectric absorption, in agreement with observation (see Sect. 1). 
``Absorbed plus unabsorbed'' modelling which was previously used predicts an intensity
decrease of typically a factor of 10 above 40 keV which is definitely not seen
and the model must be rejected as incorrect. 

Analysis of the present observation by Boirin et al. (2004b) claims that both continuum
evolution in dipping and the absorption lines could be modelled by an ionized absorber, and that
this might be able to explain {\it all} LMXB dipping sources. A consequence would be that
an extended ADC would not be required. However, this suggestion can be ruled out.
Firstly, they ignore the strong evidence from dip ingress timing 
(Church \& Ba\l uci\'nska-Church 2004) that the dominant Comptonized emission region (the ADC)
is very extended, as well as evidence cited in many papers (see Sect. 1) that one
emission component is removed gradually in dipping showing that it must be extended.
The use of an ionized absorber model means a high degree of electron scattering of the 
continuum {\it which we can rule out}. Their ionized absorber has ionization parameter
$log \,\xi$ $\sim$ 4.0 (non-dip) and 3.5 (deep dipping) so that almost all electrons are removed from ions.
Their ionized absorber column density in deep dipping is equivalent to an electron column density  
$N_{\rm e}$ of $1.8\times 10^{24}$ cm$^{-2}$. In non-dip emission, the value is $\sim$ 100
times less. Thus the increase of electron column density in dipping would cause a
decrease in X-ray continuum intensity at every energy due to Thomson
scattering by a factor of $exp -(N_{\rm e}\sigma_{\rm T})$ where $\sigma_{\rm T}$ is the Thomson
cross-section, equal to 3.2. But the high energy
spectra of several dipping LMXB in the {\it SAX} PDS instrument have already been
examined, including that of XB\th 1323-619 (BC99) in which the change in intensity in the 20 -- 50 keV
PDS band was $<$ 10$\pm$10 percent. Thus it is clear that the continuum components 
{\it are not subject to ionized absorber}, and so we can rule out the proposed model
for the dipping LMXB based on ionized absorber.

The detection of several absorption features in the non-dip spectrum of XB\th 1323-619, notably those
of Fe XXV K$_{\alpha}$ and Fe XXVI K$_{\alpha}$ allows derivation of column densities for
these species and for consideration of the site of the absorption. We use the curves of
growth for these transitions obtained by Kotani et al. (2000) in which equivalent widths
are given as a function of column density for a range of plasma temperatures between 0.1
and 1000 keV. In our case, the best-fit values are EW = 22$\pm$3 eV for Fe XXV K$_{\alpha}$,
and 27$\pm$4 eV for Fe XXVI K$_{\alpha}$. Using the curves of growth, we convert these to
column densities as shown in Table 2 for three values of the temperature used in curve 
of growth calculations $kT_{\rm CG}$: 100 keV, 45 keV and 1 keV. We will consider the two
possibilities that the absorbing plasma is formed by collisional ionization or 
photo-ionization. 

For collisional ionization, we will find the plasma thermal temperature $kT_{\rm therm}$ 
needed to give the observed ratio of column densities for Fe XXVI and Fe XXV.
Using published data on the distribution of ionization states of iron with $kT_{\rm therm}$
(Makishima 1986), we obtain $kT_{\rm therm}$ = 14 keV for $kT_{\rm CG}$ =100 keV, 
and $kT_{\rm therm}$ =100 keV for $kT_{\rm CG}$ = 1 keV. Assuming that all of the
iron in the plasma is in the forms Fe XXVII, Fe XXVI or Fe XXV, we can easily obtain
the density fraction of Fe XXV. Then, assuming solar abundances, i.e. that the ratio
H/Fe = 3.02$\times 10^4$, and using the measured column density of Fe XXV we find
the overall plasma column density $N_{\rm H}$. For example, for a curve of growth
temperature $kT_{\rm CG}$ = 100 keV, a value $N_{\rm H}$ of $9.9\times 10^{22}$ atom
cm$^{-2}$ follows. However, for a low curve of growth temperature of 1 keV,
$N_{\rm H}$ becomes very high at $2.7\times 10^{25}$ (see Table 2). This latter
possibility can be rejected since the degree of electron scattering in this plasma
would be very high with a reduction in X-ray intensity by a factor $exp(-N_{\rm e}\sigma_{\rm T})$
= $exp(-16)$. Thus X-rays would be very strongly attenuated at all energies by Thomson
scattering, and the source luminosity would be many orders of magnitude greater
than the apparent value, many times brighter than any Galactic LMXB. 

Similarly, we have obtained $N_{\rm H}$ values assuming photo-ionization. In this
case we use ion distributions as a function of the ionization parameter $\xi$
(Kallman \& McCray 1982; Makishima 1986), and find that values of $\xi$ = 2500 
and 20000 erg cm s$^{-1}$ are needed for $kT_{\rm CG}$ = 100 keV and 1 keV respectively. 
For these, the corresponding column densities $N_{\rm H}$ are $4.3\times 10^{22}$ and
$8.0\times 10^{25}$ atom cm$^{-2}$ respectively. Again, we may reject the very
high column density case. 

The problem in interpreting the curve of growth analysis is to identify the
curve of growth temperature $kT_{\rm CG}$ with a real plasma temperature.
There are several possible temperatures, firstly a thermal value
$kT_{\rm therm}$ and an ionization temperature $kT_{\rm ioniz}$. 
For collisional ionization, there would be thermal equilibrium with
$kT_{\rm CG}$ = $kT_{\rm therm}$. For photo-ionization, $kT_{\rm ioniz}$
would determine the distribution of ion types, with $kT_{\rm CG}$ = $kT_{\rm ioniz}$
which can differ from $kT_{\rm therm}$. There may also be a third temperature
as suggested by Kotani et al. (2000) characteristic of a bulk motion such as an
outflow or of turbulence. For both ionization processes, we can reject a low
curve of growth temperature of $\sim$1 keV. If we argue that $kT_{\rm CG}$
must be of the order of 100 keV to avoid $N_{\rm H}$ becoming large 
implying strong Thomson scattering, we are faced with the problem that $kT_{\rm CG}$
is inconsistent with the required $kT_{\rm therm}$ of 14 keV and
would conclude that photo-ionization must operate with $kT_{\rm CG}$ equal to
a third plasma temperature reflecting an unknown process.

However, this can be avoided for a curve of growth temperature of 45 keV, also 
shown in Table 2. For collisional ionization this requires  $kT_{\rm therm}$ = 42 keV 
and so is consistent.
$N_{\rm H}$ is $1.7\times 10^{24}$ atom cm$^{-2}$ for which the reduction by
electron scattering is $exp(-1.0)$, i.e. by three times. This is not so high
that we can reject this possibility. For photo-ionization the $N_{\rm H}$
value is also relatively low. However, both  $kT_{\rm therm}$ = 42 keV and $\xi$
= 4000 appear to be inconsistent with conditions on the outer accretion disc.
However, a temperature of $\sim$42 keV is consistent with the range of values
we have previously obtained for the electron temperature of the ADC in this source. The broadband
{\it BeppoSAX} spectrum of XB\th 1323-619 allowed a Comptonization cut-off energy
of 44.1$^{+5.1}_{-4.4}$ keV to be measured (BC99). The electron temperature
depends on the optical depth $\tau$ with $kT_{\rm e}$ = $E_{\rm CO}$ for
low $\tau$, and 3 $kT_{\rm e}$ =$E_{\rm CO}$ for high $\tau$.
This gives $kT_{\rm e}$ = 13 keV (high $\tau$) and 44 keV (low $\tau$). However, 
the measured ADC size agreed with the maximum radius for hydrostatic equilibrium
if we assume $kT$ = 44 keV (Church 2001) supporting this higher temperature.
The present result from curve of growth analysis that the ion types are consistent
with a $kT$ $\sim$ 45 keV thus suggest that the absorption lines are formed in the 
accretion disc corona, and we propose that the ADC is the site of absorption
line production in all the dipping LMXB.

Conditions in the ADC are not well understood, and the formation of the ADC has also
not been understood, with
theoretical models divided between intrinsic processes such as disc instabilities 
and external models in which irradiation of the disc by the central source leads to
formation. Our results demonstrating the very extended nature of the ADC support the irradiation
models (e.g. Jimenez-Garate, Raymond \& Liedahl et al. (2002); 
R\'o\.za\'nska \& Czerny 1996). These models show that above
the disc will be an atmosphere or boundary layer, with a hot, less dense corona above this.
The ADC temperature is less than 10 keV; however, this is not totally consistent
with X-ray spectra of sources such as XB\th 1916-053 and XB\th 1323-619 (Church 2001)
where the cut-off energies are high (81 keV and 44 keV, respectively) showing that
$kT_{\rm e}$ can be 25 keV or more.
Jimenez-Garate et al. provide calculations of proton density as a function
of height above the disc, with values of up to $10^{14}$ cm$^{-3}$ at a radius
of the outer ADC (10$^{10}$ cm; Church \& Ba\l uci\'nska-Church 2004). 
Assuming this to be constant, we estimate the column density
for a path length of 10 percent of this as  $\sim 10^{23}$ atom cm$^{-2}$, i.e. 
relatively similar to the level of $\sim 10^{24}$ atom cm$^{-2}$ implied by
the Fe absorption lines in this work. 

The formation of the ADC in illumination models takes place initially as a
hot skin which is evaporated from the accretion disc. The models assume that energy
is absorbed from the central X-ray source and then redistributed vertically to produce
a distribution of plasma density and temperature. Thus it is likely that the ion
distribution in the ADC will depend on collisional ionization at the ADC temperature, 
and is not produced by photo-ionization in the ADC itself.

In deep dipping, the Fe XXV K$_{\alpha}$ absorption line is stronger than in the
non-dip spectrum. The normalization of the line in
the dip spectra is $\sim$5 times larger than in non-dip, while the continuum at this
energy is reduced by a factor of two compared with non-dip. This clearly shows that
there is a concentration of Fe XXV ions in the bulge in the outer disc resulting in 
increased depth of the absorption line. However, the line energy
is reduced compared with non-dip showing that the ionization state is not so high
as for the absorber producing the line in non-dip emission. The reduced quality of the
dip spectra make it difficult to get accurate equivalent widths and thus accurate
ratios for the ions Fe XXVI and Fe XXV, so that a detailed numeric evaluation
of column densities is not sensible. However, for an approximate EW of 100 eV
for Fe XXV, $N_{\rm Fe XXV}$ is $\sim 10^{20}$ ion cm$^{-2}$ and the 
hydrogen column density $N_{\rm H}$ must be at least $10^{24}$ atom cm$^{-2}$.
Such values are not inconsistent with column densities in the bulge in the outer
disc as revealed from dip spectra.

Finally, we comment on the remarkable linear increase in the rate of X-ray bursting
which we show has continued unchanged over an 14 year period. This predicts that the 
gap between bursts would become zero on Jan 11, 2008. However, we do not expect
the behaviour of the source to continue unchanged till then, since as the luminosity
increases further in the band 10$^{37}$ to 10$^{38}$ erg s$^{-1}$, we expect the
source to change from a bursting source to a flaring source, since bursts are not
generally seen from high luminosity sources. Observation during this period will be 
very interesting allowing investigation of the transition. The
observation of a smooth transition between bursting and flaring would clearly
support the origin of flares as unstable nuclear burning as proposed theoretically
(e.g. Bildsten 1995).

\section*{Acknowledgments}
This work was supported in part by the Polish KBN grant KBN-1528/P03/2003/25


\begin{thebibliography}{}

\bibitem[1997]{}
Angelini L., Church M. J., Parmar A. N., Ba\l uci\'nska-Church M.,
Mineo T., 1998, A\&A, 339, L41

\bibitem[]{}
Ba\l uci\'nska-Church M., Church M. J., Oosterbroek T.,
Segreto A., Morley R., Parmar A. N.,
1999, A\&A, 349, 495

\bibitem[]{}
Ba\l uci\'nska-Church M., Barnard R., Church M. J., Smale A. P.,
2001, A\&A, 378, 847

\bibitem[]{}
Barnard R., Ba\l uci\'nska-Church M., Smale A. P., Church M. J., 2001, A\&A, 380, 494

\bibitem[]{}
Barnard R., Church M. J., Ba\l uci\'nska-Church M., 2003, A\&A, 405, 237

\bibitem[]{}
Bildsten L., 1995, ApJ, 438, 875

\bibitem[]{} Boirin L., Parmar A. N., Barret D., Paltani S., Grindlay J. E., 2004a,
A\&A, 418, 1061


\bibitem[]{} Boirin L., M\'endez M., D\'iaz Trigo M., Parmar A. N., Kaastra J. S. , 2004b,
A\&A, submitted




\bibitem[]{}
Church M. J., 2001, Adv Space Res, 28, 323

\bibitem[]{}
Church M. J., Ba\l uci\'nska-Church M., 1993, MNRAS, 260, 59

\bibitem[]{}
Church M. J., Ba\l uci\'nska-Church M., 1995, A\&A, 300, 441

\bibitem[]{}
Church M. J., Ba\l uci\'nska-Church M., 2001, A\&A, 369, 915

\bibitem[]{}
Church M. J., Ba\l uci\'nska-Church M., 2004, MNRAS, 348, 955

\bibitem[1997]{}
Church M. J., Mitsuda K., Dotani T., Ba\l uci\'nska-Church M., Inoue H., Yoshida K.
1997, ApJ, 491, 388

\bibitem[1998]{}
Church M. J., Ba\l uci\'nska-Church M., Dotani T., Asai K., 1998a, ApJ, 504, 516

\bibitem[1998]{}
Church M. J., Parmar A. N., Ba\l uci\'nska-Church M., Oosterbroek T., Dal Fiume D.,
Orlandini M., 1998b, A\&A, 338, 556

\bibitem[1986]{}
Courvoisier T. J.-L., Parmar A. N., Peacock A., Pakull M., 1986, ApJ,
309, 265

\bibitem[]{} Forman W., Jones C., Cominsky L, Julien P., Murray S.,
Peters G., Tannanbaum H., Giacconi R., 1978, ApJS, 38, 357 

\bibitem[]{}
Galloway D. K., Cumming A., Kuulkers E., Bildsten L., Chakrabarty D.,
Rothschild R. E., 2004, ApJ, 601, 466

\bibitem[]{}
Gierli\'nski M., Done C., 2002, MNRAS, 337, 1373

\bibitem[]{}
Jimenez-Garate M. A., Raymond J. C., Liedahl D. A., 2202, ApJ, 581, 1297

\bibitem[]{} Kallman T. R., McCray R., 1982, ApJS, 50, 263

\bibitem[]{} Kotani T., Ebisawa K., Dotani T., Inoue H., Nagase F., Tanaka Y.,
Ueda Y., 2000, ApJ, 539, 413

\bibitem[]{}
Oosterbroak T., Parmar A. N., Sidoli L., in't Zand J. J. M., Heise J.,
2001, A\&A, 376, 532

\bibitem[]{} Makishima K., 1986, The physics of accretion onto compact objects, Proc. of
Tenerife workshop, 1986, Lecture Notes in Physics, 266, Springer-Verlag, p249

\bibitem[1986]{}
Parmar A. N., White N.E., Giommi P., Gottwald M., 1986,
ApJ, 308, 199

\bibitem[1989]{}
Parmar A. N., Gottwald M., van der Klis M., van Paradijs J., 1989, ApJ,
338, 1024

\bibitem[]{}
R\'o\.za\'nska A., Czerny B., 1996, Acta Astron, 46, 223

\bibitem[]{}
Sidoli L., Parmar A. N., Oosterbroek T., Stella L., Verbunt F., Masetti N., Dal Fiume D.,
2001, A\&A, 368, 451

\bibitem[]{}
Smale A. P., Mason K. O., White N. E., Gottwald M., 1988,
MNRAS, 232, 647.

\bibitem[]{} 
Smale A. P., Church M. J., Ba\l uci\'nska-Church M., 2001, ApJ, 550, 962

\bibitem[]{}
Smale A. P., Church M. J., Ba\l uci\'nska-Church M., 2002, ApJ, 581, 1286

\bibitem[]{}
Stark A. A., Gammie C. F., Wilson R. W., et al., 1992, ApJS, 79, 77

\bibitem[]{}
Strohmayer T., Bildsten L., 2003, in Compact Stellar X-ray Sources, eds W. H. G. Lewin,
M. van der Klis, Cambridge

\bibitem[]{} Ueda Y., Asai K., Yamaoka K., Dotani T., Inoue H., 2001, ApJ, 556, L87

\bibitem[1992]{}
Van der Klis M., Jansen F., van Paradijs J., Stollman G., 1985,
Space Sci Rev, 30, 512

\bibitem[]{}
Walter F. M., Mason K. O., Clarke J. T., Halpern J.,
Grindlay J. E., Bowyer  S., Henry J. P., 1982, ApJ, 253, L67

\bibitem[]{}
Warwick R. S., Marshall N., Fraser G. W., et al., 1981, MNRAS, 197, 865

\bibitem[1982]{}
White N. E., Swank J. H., 1982, ApJ, 253, L61


\end{thebibliography}
\end{document}